\documentclass[conference,10pt]{IEEEtran}

\usepackage{cite}
\usepackage{amsmath,amssymb,amsfonts,amsthm}
\usepackage{algorithmic}
\usepackage{subcaption}
\usepackage{graphicx}
\usepackage{textcomp}
\usepackage{xcolor}
\usepackage{enumitem}
\usepackage{tabularx,array}

\makeatletter
\AtBeginDocument{%
  \typeout{==== The current columnsep is \the\columnsep ====}%
}
\makeatother

\newif\ifshowchanges
\showchangestrue

\ifshowchanges
  
\else
  
\fi

\begin{document}

\title{Impact of Antenna Arrays Misalignment on the Near~Field~Distance in Terahertz Communications\vspace{-4mm}}

\author{
\IEEEauthorblockN{Peng Zhang, Vitaly Petrov, Emil Bj\"ornson}
\IEEEauthorblockA{
Division of Communication Systems, KTH Royal Institute of Technology, Sweden \\
Email: \{pezhang, vitalyp, emilbjo\}@kth.se
}
}\vspace{-3mm}

\maketitle

\begin{abstract}
The extremely short wavelength of terahertz (THz) communications leads to an extended radiative near-field region, in which some canonical far-field assumptions fail. Existing near-field boundary formulations (\textit{Fraunhofer} distance) for uniform linear/planar array (ULA/UPA) configurations assume ideal alignment between transceivers, overlooking practical misalignments caused by mobility or mechanical imperfections. This paper addresses this critical gap by analyzing the impact of spatial misalignment on near-field distance calculations in THz systems. We derive exact analytical expressions and simplified approximations for the near-field boundary in both ULA--ULA and UPA--UPA configurations under arbitrary misalignment offsets. Through numerical simulations, we validate our theoretical models and quantify how misalignment reshapes the near-field region. These findings provide essential guidelines for optimizing THz system deployment in realistic scenarios.
\end{abstract}


\vspace{-2mm}
\section{Introduction}
\vspace{-1mm}
\label{sec:Intro}
Terahertz (THz, $300$\,GHz--$3$\,THz) wireless communications are considered a promising solution for enabling highly-directional ultra-high-capacity links in future networks~\cite{akyildiz2022terahertz}. However, due to a combination of (i)~extremely short wavelength of the THz signal and (ii)~the use of cm-scale antennas to achieve the desired communication range, the so-called ``radiative near-field region'' in THz communications may expand over several tens (or, ultimately, hundreds) of meters~\cite{cui2022near,proc_ieee_jornet}. Within the radiative near field, conventional far-field approximations (such as the ``plane wave assumption''~\cite{balanis2015antenna}) do not hold, requiring both near-field-specific approaches to model the signal propagation, as well as novel near-field specific communication techniques~\cite{petrov2024wavefront}. Hence, an accurate establishment of the near-field region to far-field region boundaries for a prospective deployment becomes essential.

Traditionally, the boundary between the radiative near field and the far field is established based on ensuring the maximum allowable phase difference between the planar wave model and spherical wave model does not exceed $\pi/8$. This criterion is often referred to as the \emph{``near-field distance''} or, specifically, \textit{Fraunhofer} distance~\cite{balanis2015antenna,selvan2017fraunhofer}. The expression for such near-field distance, $d_{\text{F}}$, has been originally established for a link between a point-source transmitter (i.e., an access point, AP) and a fixed-aperture receiver (i.e., a user equipment, UE). Since then, different expressions for the near-field distance have been proposed for various geometric setups involving uniform linear antenna arrays (ULAs) or uniform planar antenna arrays (UPAs). Specifically, the authors in~\cite{lu2023near} (among others) derived the near-field distance for a ULA--ULA link, while the authors in~\cite{petrov2023near} (among others) analyzed a UPA--UPA setup.

Beyond phase-based criteria, alternative definitions of the near-field boundary have been proposed from power, multiplexing, capacity, and finite-depth beamfocusing perspectives~\cite{lu2021does,lu2021communicating,cui2024near,bjornson2021primer,bohagen2009spherical,wang2014tens,jiang2005spherical}. These include the critical distance and uniform-power distance~\cite{lu2021does,lu2021communicating}, the effective Rayleigh and Björnson distances for maintaining beamforming gain~\cite{cui2024near,bjornson2021primer}, the effective multiplexing distance and eigenvalue-based thresholds for supporting spatial streams~\cite{bohagen2009spherical,wang2014tens}, and capacity-based definitions comparing spherical and planar wave models~\cite{jiang2005spherical}.

\emph{However, we observe an important gap in the prior studies.} Specifically, to the best of the authors' knowledge, all the prior studies deriving the near-field distance for the link between two ULAs or two UPAs (including~\cite{lu2023near,petrov2023near,lu2021communicating}, among others) assume perfect alignment between the antennas. Meanwhile, in realistic THz communication scenarios, both the AP and the UE will be equipped with antenna arrays that can often be misaligned due to the UE mobility. We therefore aim to address this essential gap in the present study.

A specific challenge arises here since a purely geometric approach (used in~\cite{petrov2023near,bjornson2021primer}, among others) is not directly applicable under misalignment. This is because, in such cases, the geometrically longest path between the antennas is not always the one with the largest phase shift in the received signal. Instead, the delay impact from the phase shifters must be accounted for as well, as further discussed in the paper.

\emph{To the best of the authors' knowledge, this is the first study deriving closed-form expressions for near-field distance between \underline{two misaligned antenna arrays} (ULAs and UPAs) under UE rotations.} The key contributions of this work thus~are:
\begin{itemize}
\item Closed-form exact and approximate \emph{expressions for the near-field distance} in wireless communications between two ULAs or two UPAs \emph{under various UE rotations}.
\item The near-field distance expressions get \emph{numerically elaborated} to: (i)~verify them through simulations, and (ii)~quantify the impact of antenna misalignment on the near-field to far-field boundary in THz communications.
\end{itemize}

\section{System Model}
\label{sec:Model}
\label{sec:system_model}

\begin{figure}[!t]
    \centering
    \includegraphics[width=0.85\linewidth]{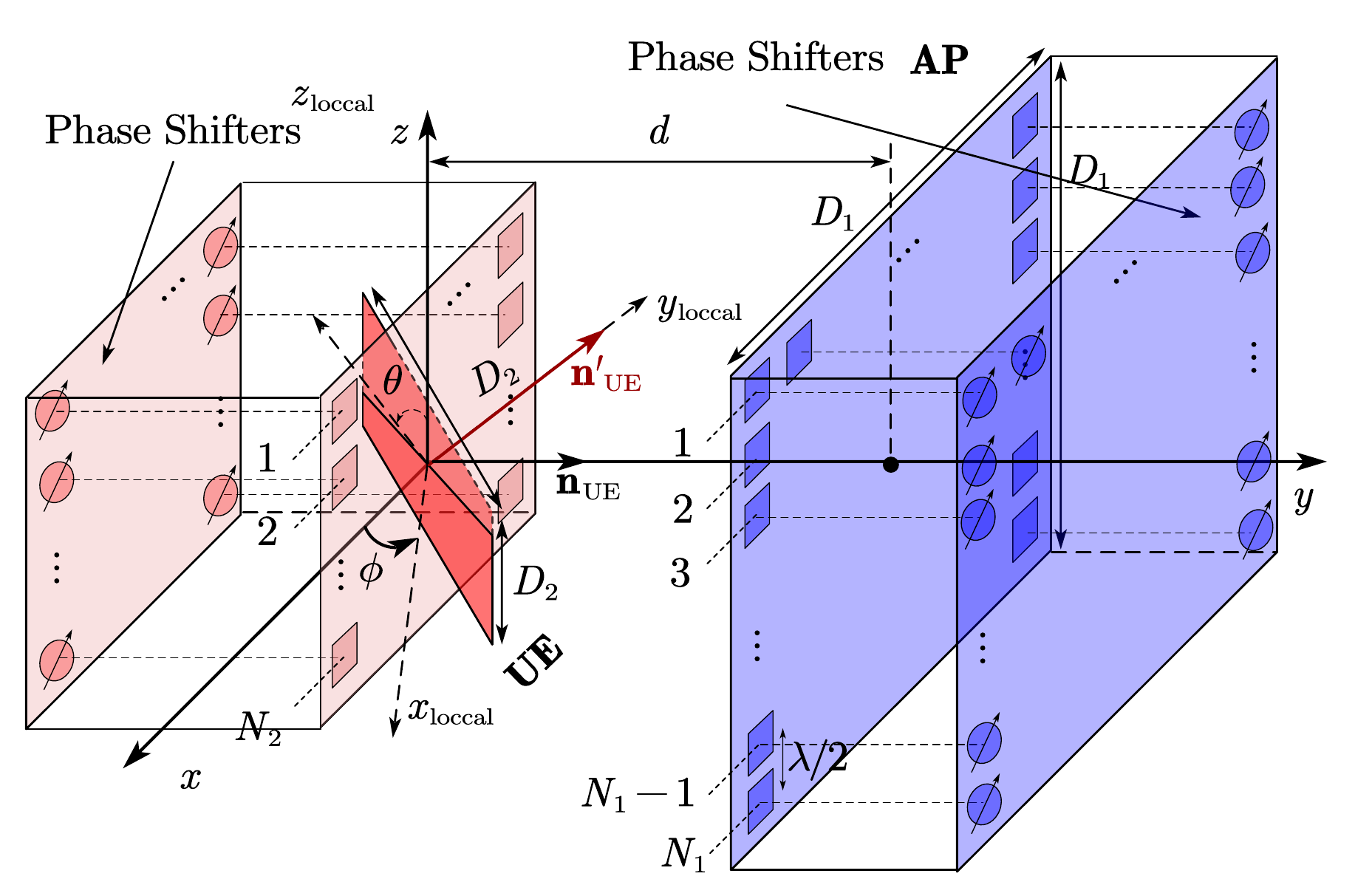}
        \vspace{-2mm}
    \caption{Modeled THz communication system with misalignment between the AP and the UE antennas caused by various UE rotations (UPA case presented, the ULA setup is similar but with an $N_1 \times 1$ AP and an $N_2 \times 1$ UE arrays).}
    \label{fig:system_model}
    \vspace{-4mm}
\end{figure}

We model a line-of-sight THz communication system (no reflected/scattered paths), as in Fig.~\ref{fig:system_model}. Both the THz AP and the THz UE are equipped with either ULAs or UPAs with the inter-element spacing of $\lambda/2$ ($\lambda = c/f$ is the wavelength in m, while $f$ is the signal frequency in Hz). The AP UPA has the physical size of $D_1 \times D_1$~m$^2$ and consists of $N_1 \times N_1$ elements, while the physical size of the UE UPA is $D_2 \times D_2$~m$^2$ and the UE array consists of $N_2 \times N_2$ elements. The UE and AP arrays are centered at the origin and on the $y$-axis, respectively, with a separation distance of $d$\,m between the centers of the antenna arrays (see Fig.~\ref{fig:system_model}).

By default, the UE lies in the $ xz$ plane, parallel to the AP. The UE can rotate in both vertical and horizontal planes (due to natural small-scale mobility~\cite{petrov_small_scale}) by $(\theta, \phi)$ angles, where $\theta, \phi \in [-\pi/2, \pi/2]$: the UE first rotates by $\theta$ around the $x$-axis, then by $\phi$ around the $z$-axis, as per the \textit{right-hand rule}. In this study, we particularly explore the angular mobility of the UE (hence, it stays on the broadside of the AP).

To rotate the UE to the spatial direction $(\theta, \phi)$, the rotation matrix is given by $\mathbf{R} (\theta,\phi) = \mathbf{R}_z(\phi) \mathbf{R}_x(\theta)$, where $\mathbf{R}_x(\theta)$ is the rotation matrix for a rotation by $\theta$ about the $x$-axis
\vspace{-0.5mm}
\begin{equation}
\mathbf{R}_x(\theta) =
\begin{bmatrix}
1 & 0 & 0 \\
0 & \cos\theta & -\sin\theta \\
0 & \sin\theta & \cos\theta
\end{bmatrix},
\end{equation}
and $\mathbf{R}_z(\phi)$ denotes the rotation matrix for a rotation by $\phi$ about the $z$-axis, which can be given by
\vspace{-0.5mm}
\begin{equation}
\mathbf{R}_z(\phi) =
\begin{bmatrix}
\cos\phi & -\sin\phi & 0 \\
\sin\phi & \cos\phi & 0 \\
0 & 0 & 1
\end{bmatrix}.
\end{equation}

\section{Near-Field Distance Calculation}
\label{sec:analysis}
We derive the near-field distance for the ULA misalignment scenario in Sec.~\ref{sec:ULA}. We later analyze the UPA misalignment scenario with the UE rotation in only one plane (Sec.~\ref{sec:UPA1}) and the UE rotation in two planes (Sec.~\ref{sec:UPA2}).

\subsection{ULA Setup with Antenna Misalignment}
\label{sec:ULA}
\begin{figure}[!b]
\vspace{-4mm}
    \centering
    \includegraphics[width=0.8\linewidth]{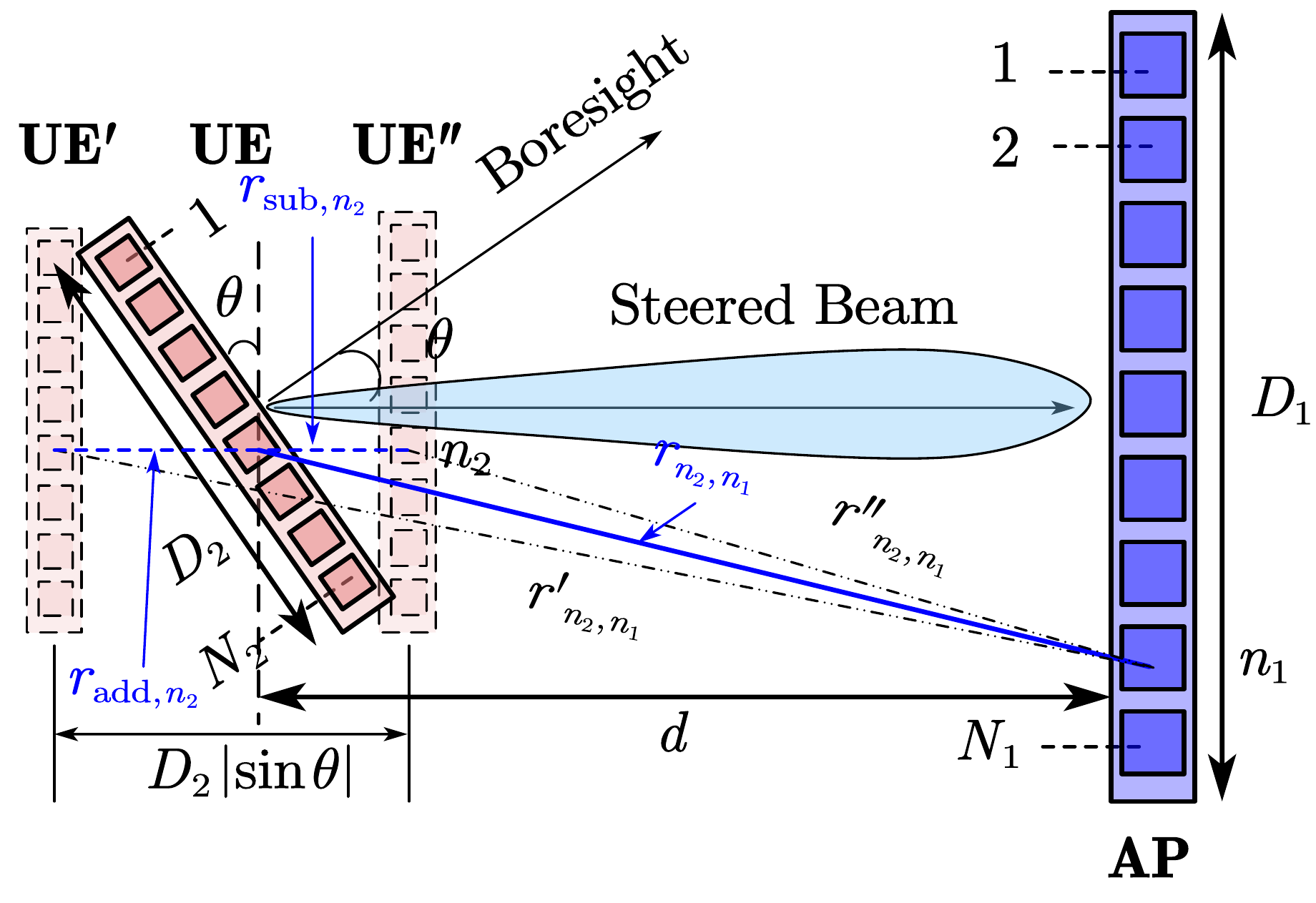}
        \vspace{-2mm}
    \caption{ULA scenario with antenna misalignment (analyzed in Sec.~\ref{sec:ULA}).}
    \label{fig:ULA_system_model}
\end{figure}

Consider the THz-ULA scenario in Fig.~\ref{fig:ULA_system_model}, where the UE misalignment is characterized by the rotation angle $\theta \in [-\pi/2, \pi/2]$. According to the geometric free space propagation assumption~\cite{sherman1962properties}, for each AP-UE antenna element pair, the phase $\phi_{n_2,n_1}$, $\forall n_1 = 1,\dots,N_1, n_2 = 1,\dots, N_2$ between the $n_1$-th element of the AP and the $n_2$-th element is
\vspace{-0.5mm}
\begin{equation}
\label{equ:system model}
\phi_{n_2,n_1}=-2\pi r_{n_2,n_1} / \lambda,
\end{equation}
where $r_{n_2,n_1}$ is the distance between the $n_1$-th antenna element of the AP and the $n_2$-th antenna element of the UE.

To steer the beam at an angle $\theta$ off the boresight (positive $\theta$ indicates rightward steering and counterclockwise UE rotation, while negative $\theta$ indicates leftward steering and clockwise rotation~\cite{delos2020phased}), the phase difference between two adjacent UE antenna elements, $\Delta\phi$, should satisfy
\vspace{-0.5mm}
\begin{equation}
\label{equ:phase compensation}
    \Delta\phi =\tilde{\phi}_{n_2+1} - \tilde{\phi}_{n_2} = \left( 2\pi / \lambda \right) \left( \lambda \sin(\theta) / 2 \right),
\end{equation}
where $\tilde{\phi}_{n_2}, \forall n_2 = {1,\dots,N_2-1}$ denotes the phase compensation required for the $n_2$-th antenna element.
Therefore, the antenna element $n_2=N_2$ requires the least phase compensation, i.e., $\tilde{\phi}_{N_2} = 0$, while the antenna element $n_2=0$ requires the most phase compensation to align with the desired beam direction. Based on the recursive relationship in \eqref{equ:phase compensation}, the phase compensation for the $n_2$-th antenna element can be given as
\vspace{-0.5mm}
\begin{equation}
    \tilde{\phi}_{n_2}=-2\pi n_2 D_2\sin (\theta) / (N_2  \lambda).
\end{equation}

Therefore, the adjusted phase between the $n_1$-th element of the AP and the $n_2$-th element of the UE can be expressed as
\vspace{-0.5mm}
\begin{equation}
\label{equ:modified phi}
    \tilde{\phi}_{n_2,n_1}=\phi_{n_2,n_1} + \tilde{\phi}_{n_2}= -2\pi \tilde{r}_{n_2,n_1} / \lambda,
\end{equation}
where
\vspace{-0.5mm}
\begin{equation}
\label{equ:adjusted r}
    \tilde{r}_{n_2,n_1}=r_{n_2,n_1}+\underset{ \triangleq r_{\mathrm{add},n_2} }{\underbrace{n_2 D_2\sin (\theta) / N_2}}.
\end{equation}

From \eqref{equ:modified phi} and \eqref{equ:adjusted r}, phase compensation can be equivalently interpreted as distance compensation. To determine the near-field distance, the maximum phase difference must not exceed $\pi/8$, which corresponds to a maximum \textit{effective distance} variation of $\lambda/16$. Thus, the near-field distance be formulated~as
\vspace{-0.5mm}
\begin{equation}\label{equ:define near-field}
    d=d_{\text{F}}^{\text{(ULA)}},\quad \mathrm{s}.\mathrm{t}. \max_{n_2,n_1} \tilde{r}_{n_2,n_1}-\min_{n_2,n_1} \tilde{r}_{n_2,n_1}= \lambda / 16.
\end{equation}

To obtain the maximum and minimum effective distances, we introduce two projected positions of the UE along the steered beam direction, denoted as $\text{UE}'$ and $\text{UE}''$, as in Fig.~\ref{fig:ULA_system_model}. Let $r'_{n_2,n_1}$ and $r''_{n_2,n_1}$ denote the distances from the $n_2$-th element of $\text{UE}'$ and $\text{UE}''$ to the $n_1$-th element of the AP, respectively. Additionally, we define $r_{\text{sub},n_2}$ to represent the ``reverse" phase compensation of the $n_2$-th element of the UE, which satisfies the following relationship with $r_{\text{add},n_2}$
\vspace{-0.5mm}
\begin{equation}
    r_{\text{add},n_2} + r_{\text{sub},n_2} \equiv D_2|\sin\theta|, \quad \forall n_2 = 1, \dots, N_2.
\end{equation}

According to the triangle inequality in Fig.~\ref{fig:ULA_system_model}, we obtain
\vspace{-0.5mm}
\begin{equation}
\label{equ:r min inequality}
    \tilde{r}_{n_2,n_1} = r_{n_2,n_1}+r_{\text{add},n_2} \geq r'_{n_2,n_1},
\end{equation}
and
\vspace{-0.5mm}
\begin{equation}
\label{equ:r max inequality}
    r_{n_2,n_1} \leq r''_{n_2,n_1}+r_{\text{sub},n_2}.
\end{equation}

By adding $r_{\text{add},n_2}$ to both sides of \eqref{equ:r max inequality}, we obtain
\vspace{-0.5mm}
\begin{equation}
\label{equ:r max inequality (new)}
    \tilde{r}_{n_2,n_1} \leq r''_{n_2,n_1}+r_{\text{sub},n_2}+r_{\text{add},n_2} = r''_{n_2,n_1} + D_2|\sin\theta|.
\end{equation}

Therefore, the range of $\tilde{r}_{n_2,n_1}$ is given by
\vspace{-0.5mm}
\begin{equation}
    r'_{n_2,n_1} \leq \tilde{r}_{n_2,n_1} \leq r''_{n_2,n_1} + D_2|\sin\theta|.
\end{equation}

According to the geometric relationships (Fig.~\ref{fig:ULA_system_model}), the maximum value of $\tilde{r}_{n_2,n_1}$ is attained when $n_1 = 1, n_2 = N_2$, hence
\vspace{-0.5mm}
\begin{equation}
        \left( \tilde{r}_{n_2,n_1} \right) _{\max}=\left( r'' _{n_2,n_1} \right) _{\max}+D_2|\sin \theta|,
\end{equation}
where
\vspace{-0.5mm}
\begin{equation}
    \left( r'' _{n_2,n_1} \right) _{\max}\!=\!\sqrt{\left( \frac{D_1}{2}\!+\!\frac{D_2}{2}\cos \theta \right) ^2\!+\!\left( d\!-\!\frac{D_2}{2}\left| \sin \theta \right| \right) ^2}.
\end{equation}
and the minimum value of $\tilde{r}_{n_2,n_1}$ is attained when $n_1 = n_2 + (D_1 - D_2)/2$. Without loss of generality, we set $n_1 = N_1/2$ and $n_2 = N_2/2$, $ \left( \tilde{r}_{n_2,n_1} \right) _{\min}$ can then be expressed as
\vspace{-0.5mm}
\begin{equation}
    \left( \tilde{r}_{n_2,n_1} \right) _{\min}=d + D_2|\sin \theta| / 2,
\end{equation}

By defining $\tilde{d} \triangleq d - D_2 |\sin \theta|/2$ and utilizing the first-order Taylor expansion $\sqrt{1+x}\approx 1+ x/2 +\mathcal{O}(x)$, we have
\vspace{-0.5mm}
\begin{equation}
\label{equ:r_max}
    \left( \tilde{r}_{n_2,n_1} \right) _{\max}\approx \tilde{d} + \left( D_1+D_2\cos \theta \right) ^2 / (8\tilde{d}) + D_2|\sin \theta|.
\end{equation}

Therefore, the equation in \eqref{equ:define near-field} can be rewritten as
\vspace{-0.5mm}
\begin{equation}\label{equ:equ_near_field_ULA}
\left( \tilde{r}_{n_2,n_1} \right) _{\max} - \left( \tilde{r}_{n_2,n_1} \right) _{\min}=\frac{\left( D_1+D_2\cos \theta \right) ^2}{8\tilde{d}}=\frac{\lambda}{16}.
\end{equation}

Substituting $\tilde{d} \triangleq d - D_2 |\sin \theta|/2$, $d=d_{\text{F}}^{\text{(ULA)}}$ into \eqref{equ:equ_near_field_ULA}, the near-field distance can be expressed as
\vspace{-0.5mm}
\begin{equation}\label{equ:ULA_near_field_dist}
    d_{\text{F}}^{\text{(ULA)}} = 2\left( D_1+D_2 \cos \theta \right) ^2 / \lambda + D_2 \left|\sin \theta\right| / 2.
\end{equation}

By symmetry, this equation also holds for the case of $\theta < 0$. Moreover, at THz frequencies, where $\lambda$ (typically $0.1$–$1$ mm) is much smaller than the physical aperture sizes $D_1$ and $D_2$ (usually several millimeters to a few centimeters), the first term in the expression dominates, while the second term, being independent of $\lambda$, can be approximated as negligible. \textbf{Thus, the near-field distance for the ULA--ULA case with the UE rotation angle $\theta$ is~approximated~as}
\vspace{-0.5mm}
\begin{equation}\label{equ:ULA near-field dist_approx}
    d_{\text{F,approx}}^{\text{(ULA)}} = 2\left( D_1+D_2 \cos \theta \right) ^2 / \lambda.
\end{equation}

Substituting $\theta = 0$, the expression reduces to the case without misalignment, given by  
$d_{\text{F}}=2(D_1+D_2)^2/\lambda$, which is consistent with the expression previously given in~\cite{lu2023near}.

\subsection{UPA Misalignment with UE Rotation over One Plane}\label{sec:UPA1}
\begin{figure}[!t]
    \centering
    \includegraphics[width=0.8\linewidth]{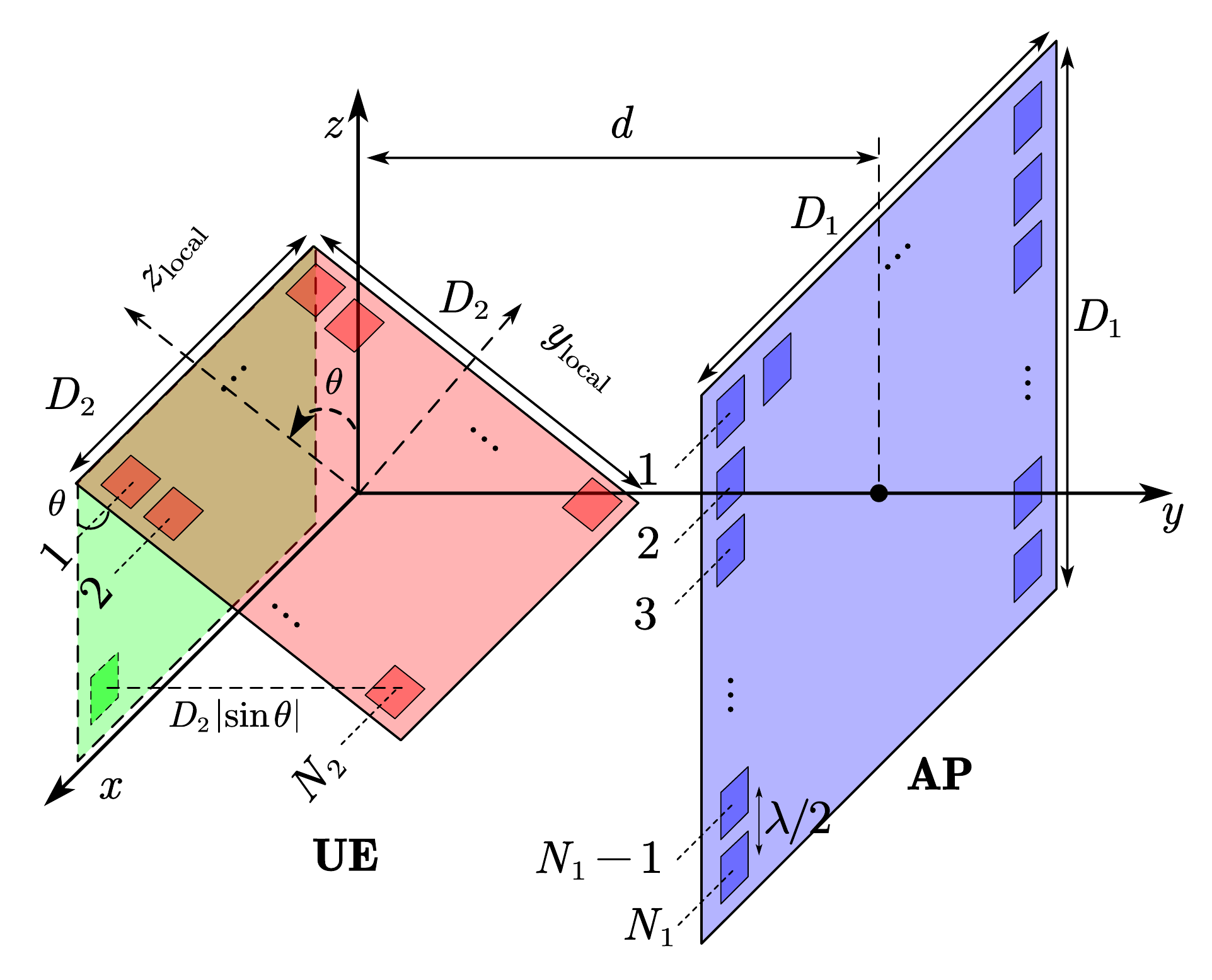}
        \vspace{-2mm}
    \caption{UPA setup with the UE rotation over one plane (Sec.~\ref{sec:UPA1}).}
    \label{fig:UPA_system_model_1}
     \vspace{-4mm}
\end{figure}
Consider the THz-UPA scenario in Fig.~\ref{fig:UPA_system_model_1}, where the UE misalignment is characterized by a rotation angle $\theta$ around the $x$-axis. Let $r_{m,n}^{i,j}$ and $\tilde{r}_{m,n}^{i,j}$, $\forall m,n=1,\dots,N_2, \; i,j=1,\dots,N_1$, denote the actual and effective distances between the $(m,n)$-th element of the UE and the $(i,j)$-th element of the AP, respectively. The near-field distance is then
\vspace{-0.5mm}
\begin{equation}\label{equ:UPA1_near_field_define}
d = d_{\text{F}}^{\text{(UPA,1)}}, \quad \mathrm{s}.\mathrm{t}. \max_{m,n,i,j} \tilde{r}_{m,n}^{i,j} - \min_{m,n,i,j} \tilde{r}_{m,n}^{i,j} = \lambda / 16.  
\end{equation}  

Inspired by Sec.~\ref{sec:ULA}, when fixing $n$ and $j$, the original problem reduces to finding the maximum \textit{effective distance} difference between $N_2$ ULA-UEs and $N_1$ ULA-APs. Define $r'^{i,j}_{m,n}$ and $r''^{i,j}_{m,n}$ as the distances from the $n$-th ULA-UE to the $j$-th ULA-AP along the steered beam direction, as in Fig.~\ref{fig:ULA_system_model}, the condition in \eqref{equ:UPA1_near_field_define} can be rewritten as
\vspace{-0.5mm}
\begin{equation}\label{equ:UPA_1_max_dist}
    \begin{split}
        &\max_{n,j} \max_{m,i} \tilde{r}_{m,n}^{i,j}-\min_{n,j} \min_{m,i} \tilde{r}_{m,n}^{i,j} 
                =\max_{n,j}\left( r''^{1,j}_{N_2,n} \right) \\
                &- \min_{n,j} \left( r'^{{N_1/2},j}_{{N_2/2},n} \right) +D_2\left| \sin \theta \right|= \lambda / 16.
    \end{split}
\end{equation}

According to the geometry in Fig.~\ref{fig:UPA_system_model_1}, the maximum value is attained when $j = N_1, n = N_2$, which is
\vspace{-0.5mm}
\begin{equation}\label{equ:UPA_1_min_dist}
     (\tilde{r}_{m,n}^{i,j})_{\max}=\sqrt{\left( \frac{D_1}{2}+\frac{D_2}{2} \right) ^2+\left( r'' _{n_2,n_1} \right) _{\max}^{2}}+D_2\left| \sin \theta \right|,
\end{equation}
and the minimum value is attained when $n = j + (D_1 - D_2)/2$. Without loss of generality, setting $j = N_1/2, n = N_2/2$, the minimum value is given by
\vspace{-0.5mm}
\begin{equation}
    (\tilde{r}_{m,n}^{i,j})_{\min}=\left( \tilde{r}_{n_2,n_1} \right) _{\min}=d + D_2 |\sin \theta| / 2.
\end{equation}

Substituting $\tilde{d} \triangleq d - D_2 |\sin \theta|/2$, \eqref{equ:UPA_1_max_dist} and \eqref{equ:UPA_1_min_dist} into the equation in \eqref{equ:UPA1_near_field_define}, we have
\vspace{-0.5mm}
\begin{equation}\label{equ:equ_near_field_UPA1}
    \left( \left( D_1+D_2 \right) ^2 + \left( D_1+D_2\cos \theta \right) ^2 \right) / (8\tilde{d}) = \lambda / 16.
\end{equation}

Substituting $d=d_{\text{F}}^{\text{(UPA,1)}}$ into \eqref{equ:equ_near_field_UPA1}, the near-field distance is
\vspace{-0.5mm}
\begin{equation}
    d_{\mathrm{F}}^{\left( \mathrm{UPA},1 \right)}=\frac{2\left( D_1+D_2 \right) ^2}{\lambda}+\frac{2\left( D_1+D_2\cos \theta \right) ^2}{\lambda}+\frac{D_2}{2}\left| \sin \theta \right|.
\end{equation}

Similarly to the path from (\ref{equ:ULA_near_field_dist}) to (\ref{equ:ULA near-field dist_approx}), \textbf{the near-field distance for the UPA--UPA case with one UE rotation angle $\theta$  is approximated as}
\vspace{-0.5mm}
\begin{equation}\label{equ:UPA1_near_field}
    d_{\mathrm{F,\text{approx}}}^{\left( \text{UPA},1 \right)} = 2 \left( \left( D_1+D_2 \right) ^2 + \left( D_1+D_2\cos \theta \right) ^2 \right) / \lambda.
\end{equation}

Compared to \eqref{equ:ULA near-field dist_approx}, \eqref{equ:UPA1_near_field} provides an intuitive interpretation of the near-field boundary for the UPA case. The first term is the Fraunhofer distance in the vertical (non-misaligned) direction, while the second term captures the horizontal component, influenced by the UE misalignment through the $\cos \theta$ factor.

\subsection{UPA Misalignment with UE Rotation over Both Planes}\label{sec:UPA2}
\begin{figure}[!t]
    \centering
    \includegraphics[width=0.8\linewidth]{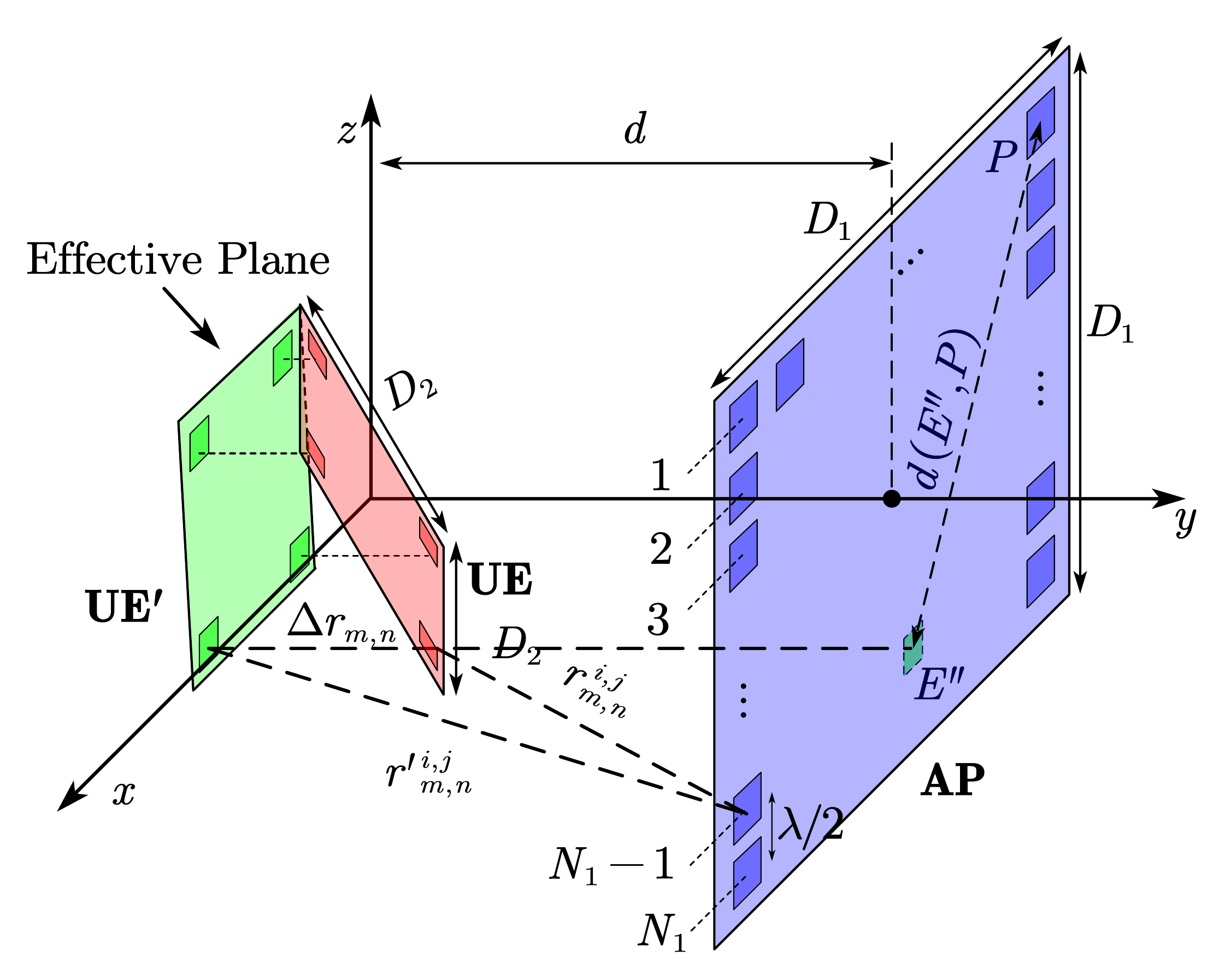}
    \caption{UPA setup with the UE rotation over both planes (Sec.~\ref{sec:UPA2}).}
    \label{fig:UPA_system_2}
    \vspace{-4mm}
\end{figure}
Consider now the \emph{full-scale THz-UPA scenario}, as in Fig.~\ref{fig:system_model}, where UE misalignment is characterized by two rotation angles $(\theta, \phi)$. Let $\mathbf{r}_{m,n}^{(\mathrm{local})}$, for $m,n = 1, \dots, N_2$, denote the initial position of the $(m,n)$-th antenna element on the UE. After applying the rotation matrix $\mathbf{R}$, the transformed global position is given by $\mathbf{r}_{m,n} = \mathbf{R} \mathbf{r}_{m,n}^{(\mathrm{local})}$. To steer the beam towards the AP which is aligned with the unit direction along the $y$-axis, i.e., $\mathbf{n}_y = (0, 1, 0)$, the required phase compensation for the $(m,n)$-th element is given by
\vspace{-0.5mm}
\begin{equation} \label{equ:std_phase_compensation}
    \Delta \phi_{m,n} = - 2\pi (\mathbf{r}_{m,n} \cdot \mathbf{n}_y) / \lambda = - 2\pi \left( \mathbf{R} \mathbf{r}_{m,n}^{(\mathrm{local})} \right) \cdot \mathbf{n}_y / \lambda .
\end{equation}  

Inspired by Sec.~\ref{sec:ULA}, phase compensation can be interpreted as shifting the points $\mathbf{r}_{m,n}$ on the UPA plane along $-\mathbf{n}_y$ by a distance $d_{m,n} = \mathbf{r}_{m,n} \cdot \mathbf{n}_y$, referred to as the compensation distance. The compensated position is given by
\vspace{-0.5mm}
\begin{equation}
\mathbf{r}'_{m,n} = \mathbf{r}_{m,n} - (\mathbf{r}_{m,n} \cdot \mathbf{n}_y) \mathbf{n}_y,
\end{equation}
which satisfies  $\mathbf{r}'_{m,n} \cdot \mathbf{n}_y = 0$. Thus, all compensated positions $\mathbf{r}'_{m,n}$ lie on the same plane with normal vector $\mathbf{n}_y$.

In practical beam steering, phase adjustments correspond to relative delays, typically defined with respect to a reference point. Hence, we introduce the effective compensation distance
\begin{equation}
\Delta r_{m,n} = d_{m,n} + |\min_{m,n} (\mathbf{r}_{m,n} \cdot \mathbf{n}_y)|.
\end{equation}

Then, the applied phase compensation can be given by $\Delta \phi' _{m,n}=-2\pi \Delta r_{m,n} / \lambda = -2\pi \left(\mathbf{r}_{m,n}\cdot \mathbf{n}_y+\Delta d\right) / \lambda,$ where $\Delta d\triangleq |\min_{m,n} (\mathbf{r}_{m,n} \cdot \mathbf{n}_y)|$. Redefining the compensated element position $\mathbf{r}'_{m,n}$ with the offset $\Delta d$, we obtain 
\begin{equation}\label{equ:compensated_r_UPA}
\mathbf{r}'_{m,n} = \mathbf{r}_{m,n} - d'_{m,n} \mathbf{n}_y = \mathbf{r}_{m,n} - (\mathbf{r}_{m,n} \cdot \mathbf{n}_y) \mathbf{n}_y - \Delta d \mathbf{n}_y.
\end{equation}


Since $\mathbf{r}_{m,n} - (\mathbf{r}_{m,n} \cdot \mathbf{n}_y) \mathbf{n}_y$ lies in the plane $\mathbf{r} \cdot \mathbf{n}_y = 0$, and $-\Delta d \mathbf{n}_y$ is a uniform shift along $\mathbf{n}_y$, the normal vector remains unchanged as $\mathbf{n}_y$. We refer to this shifted plane as the \textit{effective plane}. Inspired by Sec.~\ref{sec:ULA} and Sec.~\ref{sec:UPA1}, we approximate the solution by computing the near-field distance between the effective plane $\text{UE}'$ and the AP. A proof of the validity of this approximation is given in Appendix.

Redrawing Fig.~\ref{fig:system_model}, the updated diagram in Fig.~\ref{fig:UPA_system_2} illustrates the relationship between the effective plane $\text{UE}'$, the original UE, and the AP, along with the distances $r^{i,j}_{m,n}$, $r'^{i,j}_{m,n}$, and $\Delta r_{m,n}$. Without loss of generality, the AP element coordinates are defined as $P=(x_{\text{AP}}, d, z_{\text{AP}})$, where $x_{\text{AP}}, z_{\text{AP}} \in \left[-D_1/2 , D_1/2 \right]$, and the UE element coordinates (before rotation) are given by $(x, 0, z)$, where $x, z \in \left[-D_2/2, D_2/2 \right]$. After rotation, the UE element coordinates $E=(x_{\text{UE}}, y_{\text{UE}}, x_{\text{UE}})$ can be derived using the rotation matrix $\mathbf{R}$ as $x_{\text{UE}} = x \cos\phi + z \sin\phi \sin\theta$, $y_{\text{UE}} = x \sin\phi - z \cos\phi \sin\theta$, and $z_{\text{UE}} = z \cos\theta$.

\begin{figure*}[t]
    \centering
    \begin{subfigure}[b]{0.35\textwidth}
        \centering
        \includegraphics[width=\textwidth]{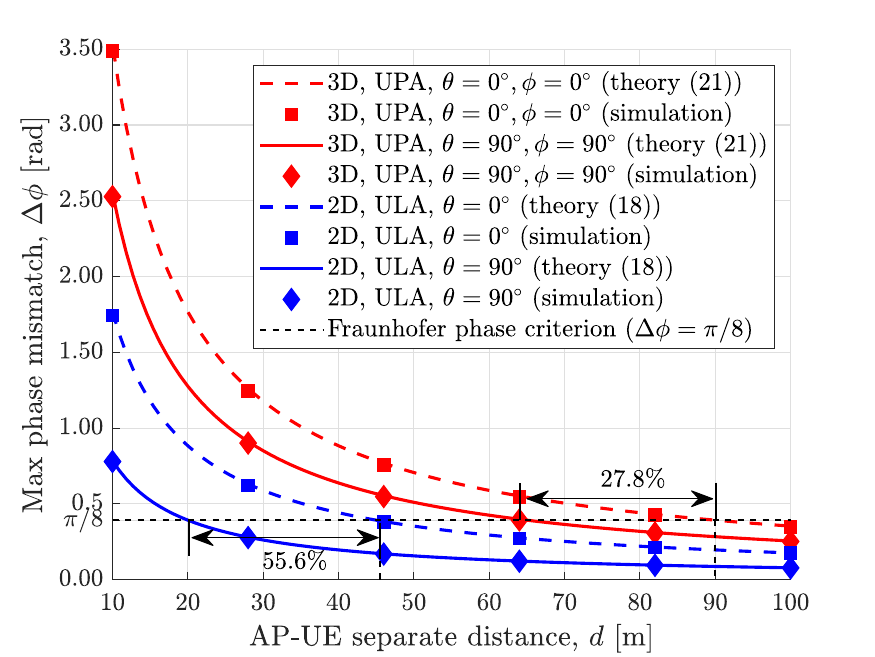}
        \caption{Maximum phase mismatch $\Delta \phi$ versus $d$.}
        \label{fig:2d_sim_phi_R}
    \end{subfigure}
    \hspace{-0.7cm}
    \begin{subfigure}[b]{0.35\textwidth}
        \centering
        \includegraphics[width=\textwidth]{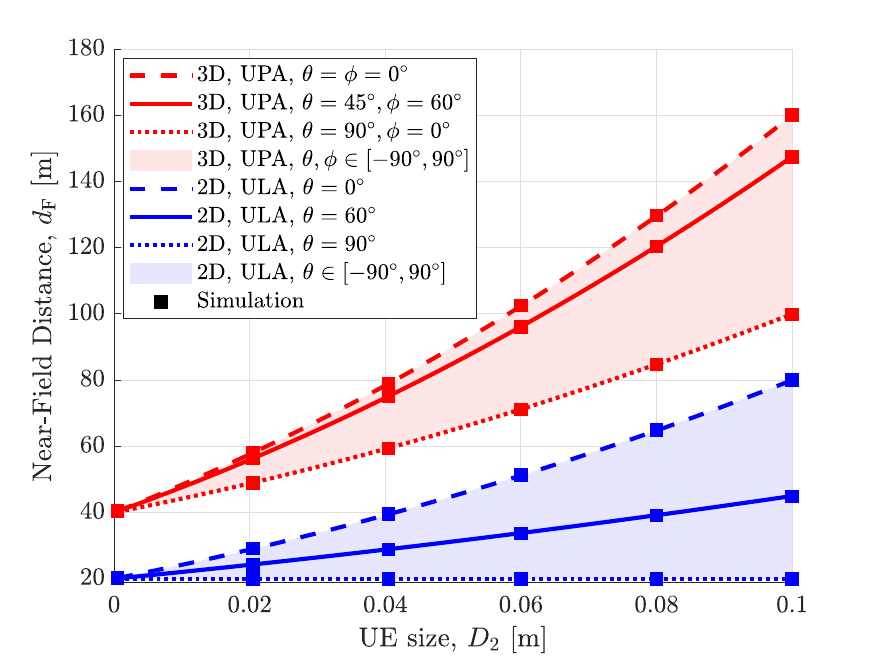}
        \caption{Near-field distance $d_{\text{F}}$ versus $D_2$.}
        \label{fig:d_F_sim}
    \end{subfigure}
    \hspace{-0.7cm}
    \begin{subfigure}[b]{0.35\textwidth}
        \centering
        \includegraphics[width=\textwidth]{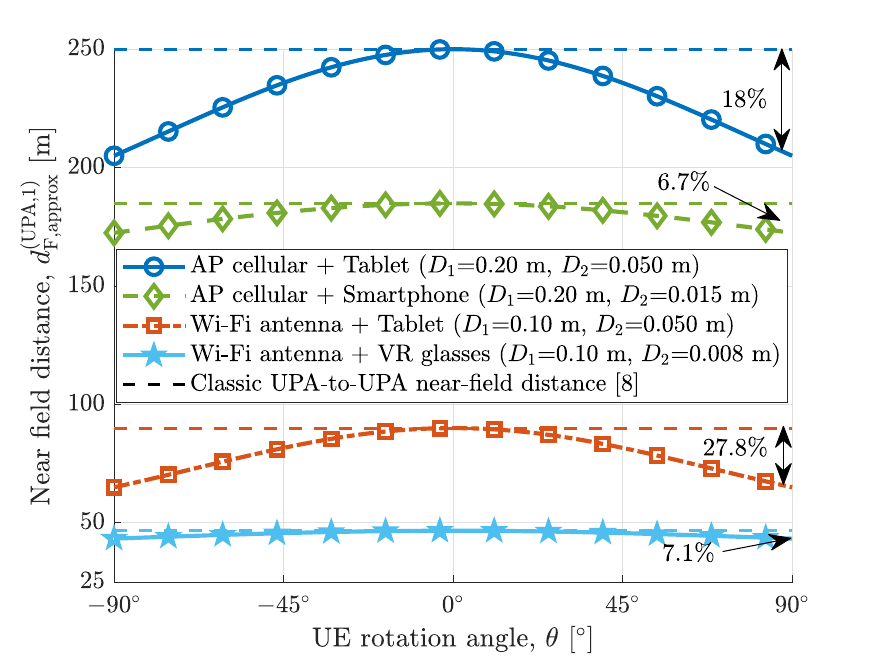}
        \caption{UPA near-field distance (one angle) versus $\theta$.}
        \label{fig:3D_sim1}
    \end{subfigure}
    \caption{The phase mismatch with $D_1=0.1\text{ m}$ and $D_2=0.05\text{ m}$ and near-field distance under different array and rotation configurations.}
    \label{fig:three_figures}
    \vspace{-5mm}
\end{figure*}

According to \eqref{equ:compensated_r_UPA}, the coordinates on the effective plane can be expressed as $E' = (x_{\text{UE}}, -\Delta d, z_{\text{UE}})$, where $\Delta d = {D_2} \left( |\cos\phi \sin\theta| + |\sin\phi| \right) / 2$. Given the parallel relationship between the $\text{UE}'$ and AP planes, the minimum distance can be expressed as
\vspace{-0.5mm}
\begin{equation}\label{equ:UPA2_min_dist}
(r'^{i,j}_{m,n})_{\min} = d + \Delta d.
\end{equation}
Furthermore, the maximum distance can be written as
\vspace{-0.5mm}
\begin{equation}
(r'^{i,j}_{m,n})_{\max} = \sqrt{(r'^{i,j}_{m,n})^2_{\min} + d_e^2},
\end{equation}
where $d_e \triangleq \max_{E'',P} d(E'',P)$ denotes the maximum distance between the projection of $\text{UE}'$ on the AP, denoted as $E'' = (x_{\text{UE}}, d, z_{\text{UE}})$, and a point $P$ on the AP. Applying the first-order Taylor expansion $\sqrt{1 + x} \approx 1 + x/2 + \mathcal{O}(x^2)$ to
\vspace{-0.5mm}
\begin{equation}\label{equ:UPA2_max_dist_approx}
(r'^{i,j}_{m,n})_{\max}  \approx r'^{i,j}_{m,n} \left( 1 + \frac{d_e^2}{2(r'^{i,j}_{m,n})^2} \right) = r'^{i,j}_{m,n} + \frac{d_e^2}{2r'^{i,j}_{m,n}}.
\end{equation}

Substituting \eqref{equ:UPA2_min_dist} and \eqref{equ:UPA2_max_dist_approx} into \eqref{equ:UPA1_near_field_define}, and further approximating with $d \gg \Delta d$, we obtain
\vspace{-0.5mm}
\begin{equation}\label{equ:near_field_dist_UPA2}
   d_{\text{F,approx}}^{\text{(UPA,2)}} =  8 d_e^2 / \lambda.
\end{equation}

According to the geometric relationships in Fig.~\ref{fig:UPA_system_2}, we define $x_{\text{UE}} = D_2 \operatorname{sgn}(\cos\phi)/2$, $z_{\text{UE}} = D_2 \operatorname{sgn}(\sin\phi \cos\theta) / 2$, $x_{\text{AP}} = -D_1 \operatorname{sgn}(x_{\text{UE}}) / 2,$ and $z_{\text{AP}} = -D_1 \operatorname{sgn}(z_{\text{UE}}) / 2$, where $\operatorname{sgn}(\cdot)$ is the sign function. Hence, the maximum distance $d(E'', P)$~is
\vspace{-0.5mm}
\begin{equation} \label{equ:UPA_2_de}
d_e = \sqrt{\left(D_1 / 2 + |x_{\text{UE}}|\right)^2 + \left( D_1 / 2 + |z_{\text{UE}}|\right)^2},
\end{equation}
where $|x_{\text{UE}}|= D_2 \left( \cos \phi + \left| \sin \phi \sin \theta \right| \right)/2$ and $|z_{\text{UE}}| = D_2\cos(\theta)/2$. Substituting \eqref{equ:UPA_2_de} into \eqref{equ:near_field_dist_UPA2}, \textbf{the near-field distance for the UPA--UPA case with two UE rotation angles ($\theta,\phi$) is approximated as}
\vspace{-0.5mm}
\begin{equation}\label{equ:near_field_UE_rotation_UPA2}
\begin{aligned}
         d_{\text{F,approx}}^{\text{(UPA,2)}} &=  2\left( D_1+D_2\left( \cos \phi +\left| \sin \phi \sin \theta \right| \right) \right) ^2 / \lambda\\
        &+2\left( D_1+D_2\cos \theta \right) ^2 / \lambda.
    \end{aligned}
\end{equation}

Letting $\phi = 0$, this expression reduces to the case with a single rotation angle, which matches the expression in \eqref{equ:UPA1_near_field}. Further setting $\theta = \phi = 0$, this formula simplifies to $d_{\text{F}} = {4(D_1 + D_2)^2}/{\lambda}$, consistent with the result given in~\cite{petrov2023near}.

\section{Numerical Results}
\label{sec:simulation}

We proceed with validating the key closed-form approximate expressions for the near-field distances $d_{\text{F}}$ derived in Sec.~\ref{sec:analysis} -- \eqref{equ:ULA near-field dist_approx}, \eqref{equ:near_field_UE_rotation_UPA2}, and \eqref{equ:UPA1_near_field} -- for the three main deployment configurations: (i)~ULA setup with the UE rotation angle $\theta$; (ii)~UPA setup with two UE rotation angles $(\theta, \phi)$; and (iii)~UPA setup with the UE rotating only in one plane with the rotation angle $\theta$ ($\phi=0$), respectively. We then systematically investigate the impact of antenna misalignment on the near-field distance for each setup. For illustration, we set the frequency to $f = 300$\,GHz (wavelength of $\lambda = 1$\,mm).


To validate our analysis, we compare the analytical results with those obtained from a custom simulator that computes the near-field distance $d_{\text{F}}$ numerically by definition~\cite{selvan2017fraunhofer,balanis2015antenna}—the smallest $d$ for which the maximum phase mismatch $\Delta \phi$ between received signal components is below $\pi/8$. For each $d$, the simulator exhaustively searches all AP–UE antenna element pairs (also accounting for phase shifter delays, see Fig.~\ref{fig:system_model}) to identify the pair producing the largest $\Delta \phi$.

We start with Fig.~\ref{fig:2d_sim_phi_R} that presents the maximum phase mismatch $\Delta \phi$ as a function of the AP-UE separation distance $d$ for both ULA and UPA configurations. A horizontal dashed line indicating $\Delta \phi = \pi/8$ criterion is provided for comparison. The intersection points between theoretical curves and this line denote the near-field distances. The analytical and simulation results in Fig.~\ref{fig:2d_sim_phi_R} show a close-to-perfect match across all the scenarios, AP-UE separation distances, and the UE rotation angles, hence verifying the accuracy of our analysis in Sec.~\ref{sec:analysis}. It can be further observed from Fig.~\ref{fig:2d_sim_phi_R} that the UE rotation decreases the near-field distance by approximately $55.6\%$ for ULA and $27.8\%$ for UPA, which highlights the importance of misalignment considerations in~our~analysis.

Further, the UPA in Fig.~\ref{fig:2d_sim_phi_R} consistently leads to larger $\Delta \phi$ than the ULA, requiring greater distances to reach the far-field region ($\Delta \phi \leq \pi/8$). Notably, in the special case of $\theta = \phi = 0^{\circ}$ for the UPA, the near-field distance reduces to $d_{\text{F}} = {4(D_1 + D_2)^2}/{\lambda}$ as in~\cite{petrov2023near}, which yields $d_{\text{F}} = 90$\,m when $D_1 = 0.1$\,m and $D_2 = 0.05$\,m. Similarly, for the ULA case with $\theta = 0^{\circ}$, the expression simplifies to $d_{\text{F}} = {2(D_1 + D_2)^2}/{\lambda}$ as in~\cite{lu2023near}, which leads to $d_{\text{F}} = 45$ m. These values further validate the correctness of the proposed analytical framework.

We proceed with Fig.~\ref{fig:d_F_sim} evaluating the impact of the UE antenna size on the near-field distance. We first again note a match between the analytical and simulation results in Fig.~\ref{fig:d_F_sim}. Therefore, for clarity, we omit the simulation results in the following figures. In Fig.~\ref{fig:d_F_sim}, we further note that both the near-field distances and their variations under misalignment increase with UE size. This trend is because the effective aperture directly affects the spatial variation in propagation paths, which determines the near-field~region.

To investigate angular effects in more detail, Fig.~\ref{fig:3D_sim1} shows the near-field distance of the UPA (under a single rotation angle) as a function of the UE rotation angle $\theta$. In this figure, we consider several realistic AP-UE dimension combinations: APs — cellular ($D_1 = 0.20$ m) and Wi-Fi antenna ($D_1 = 0.10$ m); and UEs — tablet ($D_2 = 0.05$ m), smartphone ($D_2 = 0.015$ m), and VR glasses ($D_2 = 0.008$ m). The results reveal that the near-field distance is symmetric with respect to $\theta = 0^\circ$, and decreases with greater angular misalignment. This is attributed to the reduced effective aperture with rotation, which introduces smaller propagation path variations across antenna elements. Moreover, larger UEs experience greater fluctuations in the near-field distance. For instance, under a cellular AP, the near-field distance varies by up to $18.0\%$ for a tablet, compared to only $6.70\%$ for a smartphone. Interestingly, for the same UE (tablet), a smaller AP (Wi-Fi) leads to a larger variation ($27.8\%$) than a larger cellular AP ($18.0\%$). This is due to the finer phase compensation capability of a larger cellular AP.

\begin{figure}[!t]
    \centering
    \includegraphics[width=0.85\linewidth]{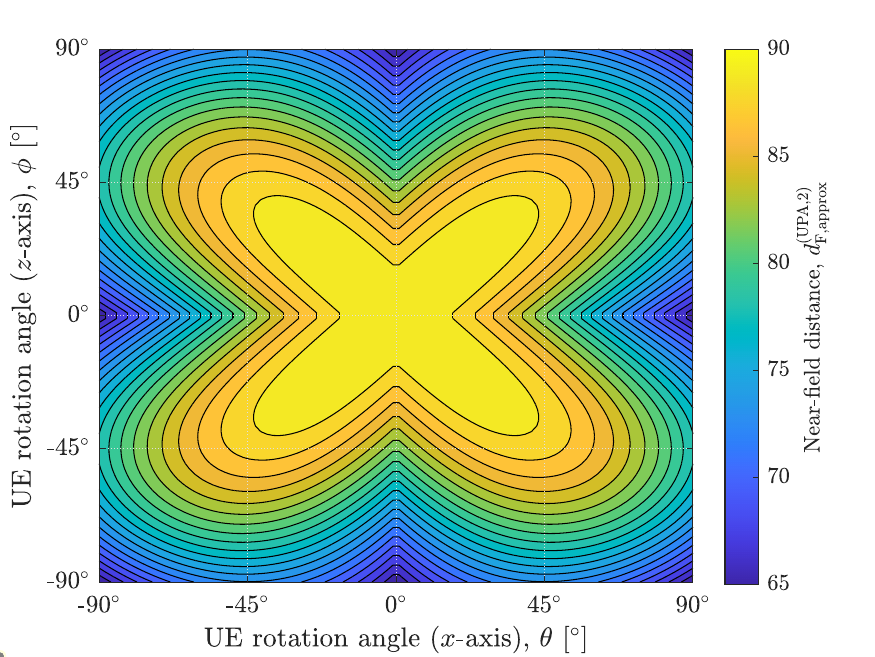}
    \vspace{-1mm}
    \caption{UPA $d_{\text{F}}$ (two angle) with $D_1=0.1\text{ m}$, and $D_2=0.05\text{ m}$.} 
    \label{fig:3D_sim1_2Dview}
    \vspace{-5mm}
\end{figure}

Finally, Fig.~\ref{fig:3D_sim1_2Dview} illustrates the joint impact of rotation angles $\theta$ and $\phi$ on the UPA near-field distance. The surface exhibits a cross-shaped pattern with contour lines showing equal near-field distances. Although one might expect the near-field distance to be determined by the projected aperture along each axis, such as $D_2 \cos\theta$ or $D_2 \cos\phi$, the observed cross pattern suggests a combined effect of $\theta$ and $\phi$. This observation highlights the nontrivial nature of AP and UE antenna misalignment in realistic mobile setups and the necessity of a full geometric treatment when evaluating near-field boundaries. Moreover, the density of the contour lines reveals that the near-field distance varies most rapidly when either $\theta$ or $\phi$ changes individually, especially around zero, while slower variation is observed along the $\theta = \phi$ direction. This is because dual-axis rotations lead to more symmetric distortions, which cause less variation in phase~mismatch.

\vspace{-1mm}
\section{Conclusion}
\label{sec:conclusion}
In this paper, closed-form exact and approximate expressions are derived for the near-field distance in THz communications with different antenna array types (ULA and UPA) and under various rotations of the UE. Unlike prior alignment-based models, our analysis incorporates realistic spatial deviations caused by the UE rotation. An accurate estimation of the near-field distance is essential to use the most appropriate and efficient (near-field versus far-field) wavefront engineering methods in next-generation wireless communication systems.

Our numerical study concludes that the array misalignment significantly affects the near-field distance in THz communications. Specifically, the UE rotation decreases the near-field distance by $\leq$$55.6\%$ for ULAs and $\leq$$27.8\%$ for UPAs, leading to up to tens of meters of mismatch with the state-of-the-art alignment-based formulas. Importantly, our analytical framework is not bounded to the THz band and can be further extended to capture the UE mobility away from the broadside of the AP, as well as other near-field to far-field boundary criteria besides the canonical $<$$\pi/8$ Fraunhofer phase mismatch.

\vspace{-1mm}
\section*{Acknowledgment}
This work has been supported by the SweWIN center (Vinnova project 2023--00572), Digital Futures at KTH, Grant 2022--04222 from the Swedish Research Council, and the Swedish Foundation for Strategic Research FFL--9 program.

\appendix
\label{appendix:proof_1}
To proof the approximation in Sec.~\ref{sec:UPA2}, let $d(\cdot, \cdot)$ denote the Euclidean distance between two points. Consider point $P$ on the AP, point $E$ on the UE, and point $E'$ on the rotated UE surface (UE$'$). Define $f(d) = d(E, P) + d(E, E') - d(E', P)$. By applying the first-order Taylor approximation $\sqrt{1 + x} \approx 1 + x/2 + \mathcal{O}(x^2)$, it follows that
\begin{equation} \label{equ:f_d_simplified}
f\left( d \right) =(\Delta d+y_{\mathrm{UE}})\frac{(x_{\mathrm{UE}}-x_{\mathrm{AP}})^2+(z_{\mathrm{UE}}-z_{\mathrm{AP}})^2}{2(d+\Delta d)(d-y_{\mathrm{UE}})}.
\end{equation}

Using the bounds $y_{\mathrm{UE}} \le \Delta d \le D_2/2$ and $(x_{\mathrm{UE}} - x_{\mathrm{AP}})^2 + (z_{\mathrm{UE}} - z_{\mathrm{AP}})^2 \le {(D_1 + D_2)^2}/{2}$, we obtain the upper bound
\begin{equation}
f(d) \le D_2 (D_1 + D_2)^2 / (4d^2) = o\left( 1 / d^2 \right),
\end{equation}
which means that $f(d)$ is a higher-order infinitesimal of $1/d$ and can be neglected in near-field distance calculations. This implies that $d(E',P)$ can be used as an approximation for $d(E,P) + d(E,E')$ in the calculation.

\bibliographystyle{IEEEtran}
\bibliography{references}

\end{document}